\documentclass{aa}  
\usepackage{graphicx}
\usepackage{color}
\usepackage{amsmath}	
\usepackage{txfonts}
\usepackage{hyperref}

\usepackage{notes2bib}

\hypersetup{colorlinks=true,linkcolor=[rgb]{1.,0.2,0.2},citecolor=[rgb]{0.1,0.4,1.},filecolor=[rgb]{0.7,0.2,0.2},urlcolor=[rgb]{0.7,0.2,0.2}}
\definecolor{darkspringgreen}{rgb}{0.09, 0.45, 0.27}
\definecolor{amber(sae/ece)}{rgb}{1.0, 0.49, 0.0}

\begin{document} 
   \title{VEXAS: the VISTA EXtension to Auxiliary Surveys}
   \subtitle{Data Release 1: the Southern Galactic Hemisphere}
   \author{C. Spiniello\inst{1,2}\fnmsep\thanks{email:chiara.spiniello@inaf.it}
          \and
          A. Agnello\inst{3}\thanks{ORCID 0000-0001-9775-0331}
          }
   \institute{INAF - Osservatorio Astronomico di Capodimonte, Salita Moiariello, 16, I-80131 Napoli, Italy\\
         \and
             European Southern Observatory, Karl-Schwarschild-Str. 2, 85748 Garching, Germany\\
         \and
             DARK, Niels Bohr Institute, Copenhagen University,  Lyngbyvej 2, 2100 Copenhagen, Denmark\\
}

   \date{Accepted, August 29, 2019}
 \abstract
  {We present the first public data release of the VISTA EXtension to Auxiliary Surveys (VEXAS), comprising of 9 cross-matched multi-wavelength photometric catalogues where each object has a match in at least two surveys.}
  {We aim at a spatial coverage as uniform as possible in the multi-wavelength sky, with the purpose of providing the astronomical community with reference magnitudes and colours for various scientific uses, including: object classification (e.g. quasars, galaxies, and stars; high-z galaxies, white dwarfs, etc.); photometric redshifts of large galaxy samples; searches of exotic objects such as, for example, extremely red objects and lensed quasars.}
  {We have cross-matched the wide-field VISTA catalogues (the VISTA Hemisphere Survey and the VISTA Kilo Degree Infrared Galaxy Survey) with the AllWISE mid-infrared Survey, requiring that a match exists within $10\arcsec$. We have further matched this table with X-Ray and radio data (ROSAT, XMM, SUMSS). We also performed a second cross-match between VISTA and AllWISE, with a smaller matching radius ($3\arcsec$), including WISE magnitudes. We have then cross-matched this resulting table ($\approx138\times10^6$ objects) with three photometric wide-sky optical deep surveys (DES, SkyMapper, PanSTARRS). We finally include matches to objects with spectroscopic follow-up by the SDSS and 6dFGS.}
  {To demonstrate the power of all-sky multi-wavelength cross-match tables, we show two examples of scientific applications of VEXAS, in particular using the publicly released tables to discover strong gravitational lenses (beyond the reach of previous searches), and to build a statistically large sample of extremely red objects.}
  {The VEXAS catalogue is currently the widest and deepest, public, optical-to-IR photometric and spectroscopic database in the Southern Hemisphere.} 
 

   \maketitle
%
%________________________________________________________________

\section{Introduction}
Wide-field Sky surveys have become a fundamental tool for observational cosmology, e.g. to explore the  large  scale  structure  and  the  role of dark matter and dark energy in the evolution of the Universe (e.g. KiDS via weak lensing, \citealt{Kuijken15,Hildebrandt17}; ATLAS, \citealt{Shanks15}; DES, \citealt{des16}) or the discovery of primordial quasars (out to the re-ionization epoch,  \citealt{Venemans13,Carnall15,Reed15, Banados16, Chehade18}), as well as for extragalactic astrophysics, e.g. the formation and evolution of galaxies, including the Milky Way, or the distribution of dark matter around galaxies (e.g. SDSS, \citealt{SDSS_DR14}; The Two Micron All Sky Survey, 2MASS, \citealt{Skrutskie06}; The Cosmic Evolution Survey, \citealt{Scoville07}; The Gaia Mission, \citealt{Gaia16}; The Galaxy And Mass Assembly, GAMA, \citealt{Driver11, Liske15}).

Over the last decade, astronomical surveys provided new insights on the physics of objects at all scales, from giants Early-Type Galaxies (ETGs) to faint and compact stellar systems, and at all distances, from the structure and dynamics of our own galaxy to high redshifts quasars. Tracing the mass assembly of the big structures in our Universe, mapping the evolution of galaxy size-mass relation, detecting and studying the intracluster light components, understanding the transient phenomena and the time-domain in the Universe, studying early cosmic times, are only few of the %enormous number of 
astrophysical and cosmological questions that have been tackled by wide-field sky surveys.

Given the value of wide and deep photometric databases, calibrated and cross-matched tables with homogeneous spatial and wavelength coverage are fundamental.% 
This is indeed the purpose of the VISTA EXtension to Auxiliary Surveys (VEXAS) project, which we present here. 

In this first data release, covering the Southern Galactic Hemisphere, we cross-match the two main extragalactic surveys on the Visible and Infrared Survey Telescope for Astronomy (VISTA, \citealt{Emerson06}), with many of the most successful wide-sky photometric surveys in the optical (the Dark Energy Survey, \citealt{Abbott18}; the Panoramic Survey Telescope \& Rapid Response System DR1, PanSTARRS1, \citealt{Chambers16}; and SkyMapper Southern Sky Survey, \citealt{Wolf18}),
in the infrared (the Wide-Infrared Survey Explorer, WISE, \citealt{Wright10}), as well as in the X-ray (ROSAT All Sky Survey, \citealt{Rosat_I, Rosat_II}; The XMM-Newton Serendipitous Survey, \citealt{XMM}), and in the radio domain (SUMSS, \citealt{Sumss_I}). 
The core requirement is a reliable photometry in more than one band. This condition, together with the detection in at least two surveys (via cross-match) should minimize, if not completely eliminate, the number of spurious detections in the final catalogues.
Finally, we also identify all the objects retrieved by the VISTA catalogues that have a spectroscopic redshift estimate from the Sloan Digital Sky Survey (SDSS, DR14, \citealt{SDSS_DR14}) and/or from the 6dF Galaxy Survey (6dFGS, \citealt{6dfgs_I}). 

All the resulting tables are validated and publicly released through the ESO Phase 3\footnote{\url{https://www.eso.org/sci/observing/phase3.html}. While processing the phase 3 documentation and format, we release the table via a temporary repository (\href{https://drive.google.com/drive/folders/18IjYlkKrvEB2AEcj6UZiHTOUKbijULpf?usp=sharing}{here}). }.

The paper is organized as follow. In Section~\ref{sec:VISTA} we provide a brief description of the VISTA Surveys (VHS and VIKING), which are the starting point of the matches, and on the criteria %we imposed to build up
leading to our VEXAS final table of 198'608'360 objects in the Southern Hemisphere. In Sections~\ref{sec:WISE} and ~\ref{sec:optical} we introduce the infrared and optical public surveys and the multi-wavelength cross-match. For each cross-match, we provide the total number of retrieved objects and main characteristics of the final table.
In Section~\ref{sec:spec_xmatches} we focus on the spectroscopic cross-matches, also giving final relative numbers of matched objects. 
In Section~\ref{sec:otherwave} we describe the matches with surveys covering larger wavelength bands, namely X-ray and radio. 
In Section~\ref{sec:science} we provide two examples of interesting scientific cases where the VEXAS tables are clearly helpful: the identification of extremely red objects, and the search for strong gravitationally lensed quasars.
Finally, we conclude in Section~\ref{sec:conclusions} and discuss possible developments for future data releases. 

We always keep the native magnitude system of each Survey unless a direct comparison is required between magnitudes given in different systems \footnote{When comparing magnitudes from different surveys, we adopt the conversions given here \url{http://casu.ast.cam.ac.uk/surveys-projects/wfcam/technical/filter-set}.}.

%__________________________________________________________________

\section{The VISTA Surveys}
\label{sec:VISTA}
VISTA, the Visible and Infrared Survey Telescope for Astronomy \citep{Emerson06}, hosted by the Paranal ESO Observatory and working at near-infrared wavelengths, has a primary mirror of 4.1 meters and was conceived to map the sky systematically in a "survey" manner. Most of the observing time was allocated to six public surveys developed by a consortium of 18 universities in the United Kingdom\footnote{VISTA is an in-kind contribution to ESO as part of the UK accession agreement}. 

%Some of this surveys focus on small patches of the sky with the goal of reaching high image depth to observe extremely faint objects (e.g. UltraVISTA, \citealt{McCracken12}), some other surveys instead aim at obtaining a uniform coverage of the entire sky. 
%Since this is indeed also the purpose of this paper, we focus here on the two VISTA Public Surveys belonging to this latter group: The VISTA Hemisphere Survey (VHS, \citealt{McMahon12}) and the VISTA Kilo Degree Infrared Galaxy Survey (VIKING, \citealt{Sutherland12}). 

For the purposes of this paper, i.e. a homogeneous and wide database for multi-wavelength studies, we focus on two VISTA Public Surveys with the largest footprints: The VISTA Hemisphere Survey (VHS, \citealt{McMahon12}) and the VISTA Kilo Degree Infrared Galaxy Survey (VIKING, \citealt{Sutherland12}). 
Both surveys share the same reduction and processing. Astrometric and photometric calibrations are based on 2MASS \citep{Skrutskie06} stars: a set of colour equations is used to predict VISTA native magnitudes from the observed 2MASS $J,H,Ks$ colours \citep{hod09}. 

\subsection{The VISTA Hemisphere Survey}
The VISTA Hemisphere Survey (VHS, \citealt{McMahon12}) is aimed at covering the Southern celestial hemisphere to $\approx30$~times deeper limits than 2MASS in at least two near-IR bands ($J$ and $Ks$). With the advent of different optical surveys covering some areas of the Southern sky, VHS would provide complementary data to the $grizY$ broad-band photometry from the Dark Energy Survey (DES, \citealt{Abbott18}) in the South Galactic Cap, and $ugriz$ from VST-ATLAS survey \citep{Shanks15}. 
The target depth (in Vega magnitudes) is $J=$20.6, $H=$19.8, $Ks=$18.5 from 120-second integration on the DES overlap, and somewhat shallower ($J=$20.2, $Ks=$ 18.1; 60 seconds per band) on the VST-ATLAS overlap.

\subsection{The VISTA-Kilo-Degree Infrared Galaxy Survey}
The VISTA-Kilo-Degree Infrared Galaxy Survey (VIKING, \citealt{Sutherland12,Edge13}) covers 1500 square degrees overlapping the Kilo Degree Survey (KiDS, \citealt{Kuijken15}) for the optical counterpart, ensuring a uniform coverage on $\sim1350$ deg$^{2}$ with intermediate depth in nine bands ($grizYJHKs$, with limiting magnitude J$=21$ in the Vega system), enabling accurate photometric redshift measurements for weak-lensing studies \citep{Hildebrandt17}. 
Among the science goals of the KiDS and VIKING, there are:  the characterization of galaxy clusters up to $z\sim1$ \citep{Maturi19}, the search for high-$z$ objects \citep{Venemans15} and new strong gravitational lenses \citep{Spiniello18, Petrillo19, Khramtsov19}, and the study of galaxy structural parameters and stellar masses for a statistically large sample of galaxies \citep{Roy18}. 

The sky coverage is at high galactic latitudes, and includes two main stripes of $\sim 70 \times 10$ degrees each: one in the South Galactic cap near dec.$\sim-30$~deg, and one near dec.$\sim0$~deg in the North galactic cap. By design, the VIKING and VHS footprints are complementary (see Fig.~\ref{fig:ra_dec_sky}). 

\subsection{Querying the VISTA Surveys}
\label{sec:query_VISTA}
\begin{figure*}
\includegraphics[width=18cm]{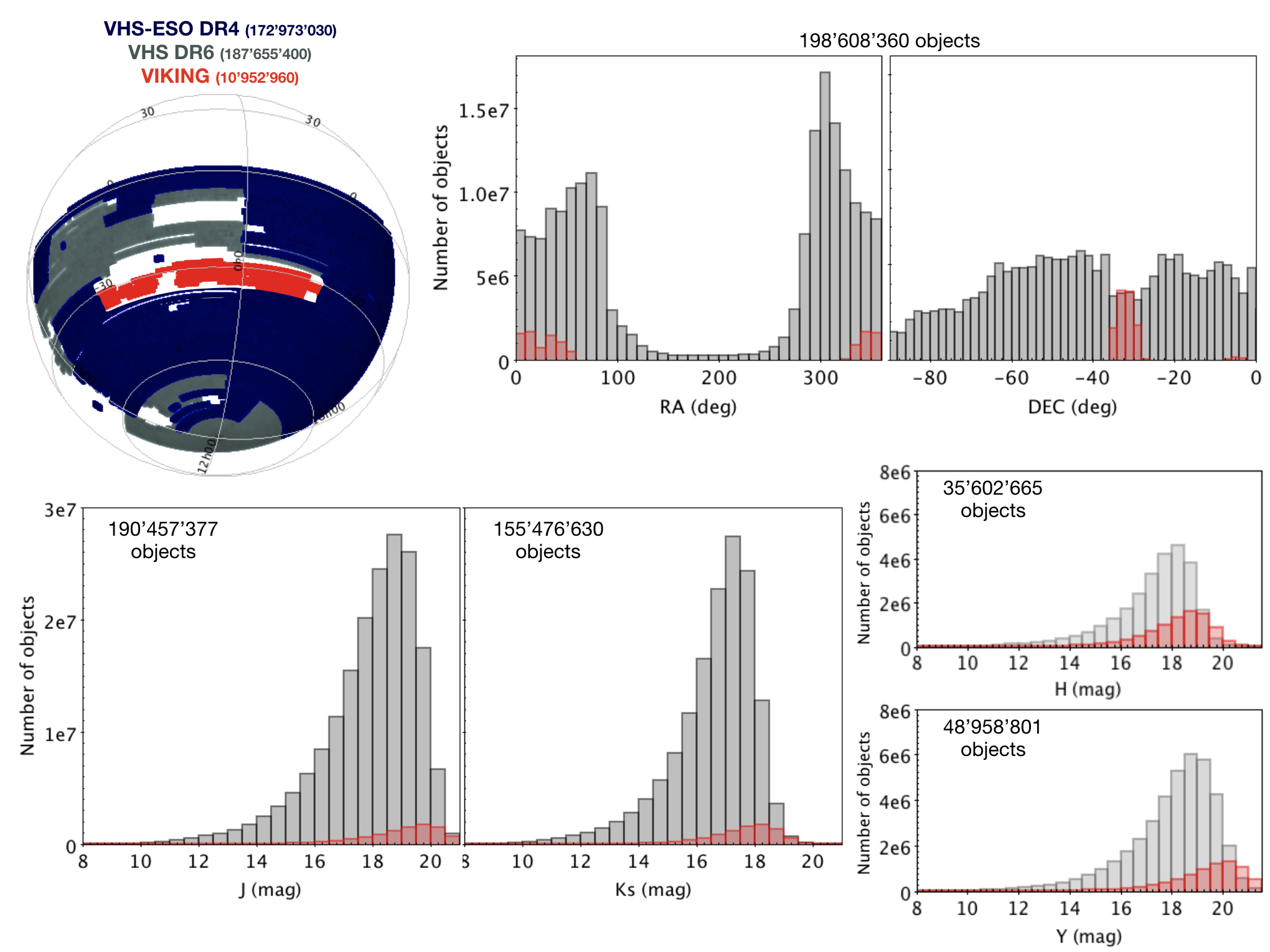}
\caption{{\sl Left:} Sky view of the VISTA footprint (VHS in grey and VIKING in red) compared to the VHS-ESO DR4 (blue). {\sl Right:} histograms of r.a. and dec. for the VISTA Surveys that we used for this VEXAS-DR1. The total number of objects in the catalogue is reported on top of each histogram. {\sl Bottom:} Distribution of $Ks,$ $J,$ $H$ and $Y$ magnitudes of the input VISTA table, with number of objects within the magnitude ranges of Eq.\ref{eq:eq_VISTA} given on the top-left of each panel. The VSH has a very large coverage in $J$ and $Ks,$ but only observed $H$ and $Y$ in few fields.}
\label{fig:ra_dec_sky}
\end{figure*}

We queried VIKING and the VHS-DR6 from the VISTA Science Archive website\footnote{\url{http://horus.roe.ac.uk/vsa/}} directly. We caution the reader that the VHS-DR6 does not match the latest release from the ESO archive, as visible from Figure~\ref{fig:ra_dec_sky}. 

To ensure that only objects with reliable photometry in at least one band are retrieved, we imposed the following criteria on the Petrosian magnitudes and their uncertainties: 

\begin{equation}
\begin{split}
\, & (\, \Delta \rm Ks < 0.3 \,  \rm AND \, 8 < \rm Ks )\ & \ \rm OR \\
\, & (\, \Delta \rm H < 0.3 \, \rm AND \, 8 < \rm H )\ & \ \rm OR \\
\, & (\, \Delta \rm J < 0.3 \, \rm AND \, 8 < \rm J)\ & 
\end{split}
\label{eq:eq_VISTA}
\end{equation}

Moreover, for this VEXAS-DR1 we only consider objects below the galactic plane, $b<-20.$ This is the area where wide-field weak-lensing cosmological experiments overlap, it covers a hemisphere with smaller previous coverage, and also includes the well-studied Stripe-82 area of the Sloan Digital Sky Survey. Comparable operations at $b>20$ are planned for the second release of VEXAS. 

\begin{figure}
\includegraphics[width=0.99\columnwidth]{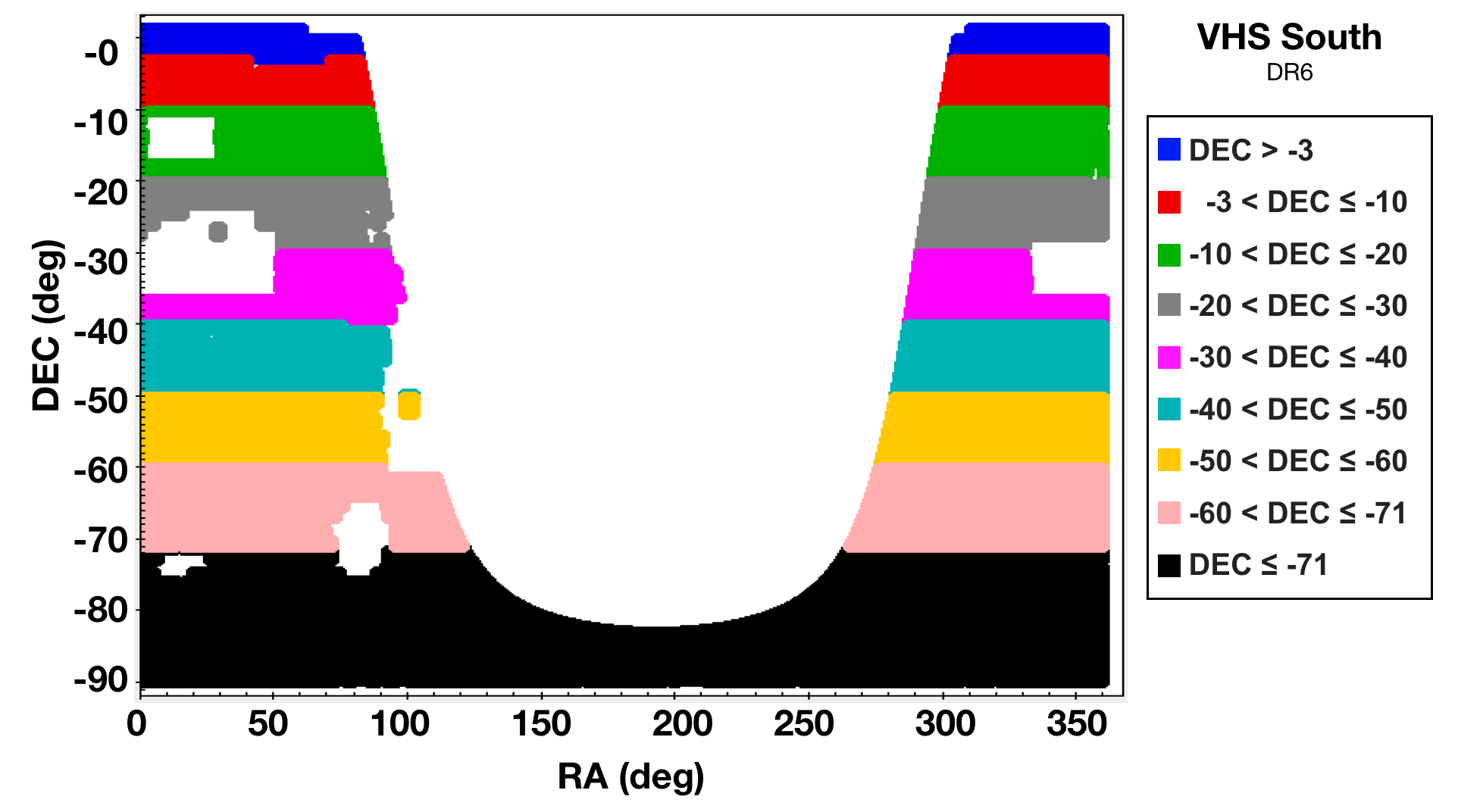}
\caption{Final coverage (r.a., dec.) of the products retrieved from the VHS DR6 for this release of VEXAS. The colours indicate different declination slices that we built to overcome the TAP limitations on size/entries. The number of objects for each slice is reported in Table~\ref{tab:ra_dec_objects}. }
\label{fig:ra_dec_slice}
\end{figure}

Given the TAP limitations on the maximum file size and maximum number of objects per query, we had to split VHS in several declination slices.  
We show the slices in Figure~\ref{fig:ra_dec_slice} and report the number of objects retrieved for each slice in Table~\ref{tab:ra_dec_objects}. 
The whole VIKING-DR3 could instead be retrieved with a single ADQL job. 

\begin{table}
\caption{Number of objects in each declination slice, plotted in Figure~\ref{fig:ra_dec_slice}.}
\label{tab:ra_dec_objects}
\begin{center}
\begin{tabular}{cr}
\hline
{\bf Declination}		&	{\bf Objects} \\
\hline

$+3 < Dec \le -3$          &    9'439'341 \\
$-3 < Dec \le -10$         &    15'829'509 \\
$-10 < Dec \le -20$        &    23'321'667 \\
$-20 < Dec \le -30$        &    21'954'249 \\
$-30 < Dec \le -40$        &    20'453'315 \\
$-40 < Dec \le -50$        &    27'090'911 \\
$-50 < Dec \le -60$        &    25'319'673 \\
$-60 < Dec \le -71$        &    22'406'811 \\
$-71 < Dec$                &    21'839'924 \\
\hline
\end{tabular}
% \begin{tablenotes}
%  \item {\sl Col.~} 
%  \item {\sl Col.~} 
%  \end{tablenotes}
  \end{center}
\end{table}

For the VEXAS-DR1, both the VIKING and VHS queries included a \texttt{JOIN} command to the cross-neighbour AllWISE table, provided by the VISTA Science Archive\footnote{\url{http://horus.roe.ac.uk/vsa/www/vsa_browser.html}} (VSA, \citealt{Cross12}), containing all the AllWISE sources within $10.0\arcsec$ of each VHS source.
No further requirements were made on the AllWISE entries, except that they exist. 

Finally, after querying VIKING and VHS separately, we concatenated the tables using the \textsc{Tool for OPerations on Catalogues And Tables (TOPCAT, \citealt{Taylor05})}\footnote{\url{http://www.star.bris.ac.uk/~mbt/topcat/}}. 
The final number of objects retrieved with these criteria is 198'608'360, of which 187'655'440 from VHS+AllWISE and 10'952'920 from VIKING+AllWISE. We report in Table~\ref{tab:numb_objects} the number of objects retrieved for each of the cross-matches we carried on starting from the VEXAS table described above. 

\begin{table}
\caption{Number of objects in VEXAS, from VISTA and subsequent cross-matches. {The optical cross-matched tables are built from the VEXAS-AllWISE3 (matching radius of 3$\arcsec$). The spectroscopic table uses SDSS and 6dFGS and it is also matched with VEXAS-AllWISE3. The X-ray table used the 2RX tables by \citet{Salvato18} and the 21cm table is obtained from SUMSS. In this last two cases, the match with AllWISE is within a 10$\arcsec$ matching radius, given the poorer resolution of X-ray and 21cm surveys.}}
\label{tab:numb_objects}
\begin{center}
\begin{tabular}{lr}
\hline
{\bf Cat.~Table}	&	{\bf Objects} \\
\hline
VEXAS & 198'608'360\\
VEXAS-AllWISE3 & 126'372'293\\
VEXAS-DESW &     37'615'619 \\
VEXAS-PS1W &     24'693'386 \\ 
VEXAS-SMW  &     20'331'041 \\
VEXAS-EXTRAGAL &  5'742'300\\
VEXAS-SPEC &       347'076 \\
VEXAS-21cm &        77'338 \\
VEXAS-Xray &         3'049\\
\hline

\end{tabular}
\end{center}
\end{table}

\section{Infrared photometric Xmatches: VEXAS-AllWISE}
\label{sec:WISE}
The Wide-Infrared Survey Explorer (WISE, \citealt{Wright10}) NASA mission provided all-sky astrometry and photometry in four mid-infrared bandpasses centred on 3.4$\mu\mathrm{m}$ (hereafter W1),  4.6$\mu\mathrm{m}$ (hereafter W2), 12$\mu\mathrm{m}$ (hereafter W3) and 22$\mu\mathrm{m}$ (hereafter W4).  

The NeoWISE program \citep{Mainzer11} extended the original WISE mission after its cryogenic phase. Combining these two complete sky coverage epochs, the AllWISE data have enhanced photometric sensitivity and accuracy, and improved astrometric precision compared to the previous WISE All-Sky Data Release. 
The WISE science goals include the search and characterization of the very luminous red galaxies, the search for nearby coolest stars, and the study of the nature of asteroids and comets. 

The AllWISE Source Catalogue (\citealt{Cutri13}) used for this VEXAS release includes $\approx$747 million objects over the whole sky, with median angular resolution of 6.1, 6.4, 6.5, and 12.0 arcseconds respectively for the four bands.  

Further to the \texttt{JOIN} with the cross-neighbour AllWISE table directly from the VISTA Science Archive, we perform a further match with a smaller matching radius of $3\arcsec,$ retrieving also the WISE magnitudes and their associated errors. The choice of $3\arcsec$ is a compromise between the resolution of WISE ($\approx6\arcsec$) and spurious cross-matches between different sources or with ghosts and artefacts.  
The VEXAS-AllWISE3 table (`3' indicates the matching radius) comprises 126'372'293 objects with photometry in at least one VISTA and in the WISE bands ($W1$ to $W4,$ with errors). For completeness, we also provide the 2MASS \citep{Skrutskie06} IDs and magnitudes in $JHKs$ bands, whenever they are available. 

{Throughout the paper, and for all the further cross-matches, we use the VEXAS-AllWISE3 table as a starting point -- unless otherwise specified. The  \texttt{JOIN} cross-match with AllWISE3 ensures a large wavelength final baseline for our catalogues, and at the same time limits (or completely eliminates) spurious detections, given that each object has at least three independent matches from three different surveys. }

\begin{figure*}
\includegraphics[width=18cm]{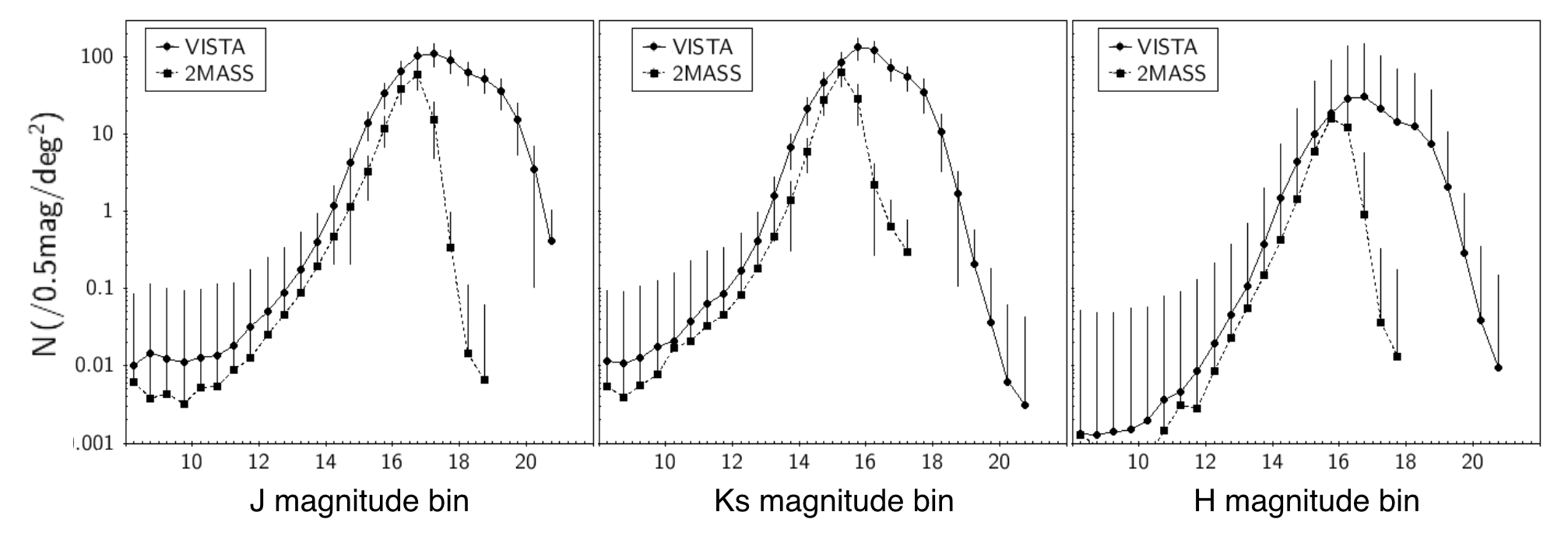}
\caption{Luminosity functions in $J$ (left), $Ks$ (middle) and $H$ (right) bands from VISTA  (circles, continuous line) and 2MASS (squared, dotted lines) in bins of 0.5 magnitudes (in the Vega system).}
\label{fig:lum_function}
\end{figure*}

The luminosity functions of VISTA-AllWISE objects are shown in Figure~\ref{fig:lum_function}. For the sake of comparison, we also show the 2MASS luminosity functions in the same ranges. This release of VISTA-AllWISE objects extends the 2MASS depth by 2-3~magnitudes. This is consistent with the target depths of the VHS and VIKING. The main limitation at the faint end is given by the requirement that the magnitude errors be $\leq0.3$mag. 
The luminosity functions corresponding to $\leq0.1$mag errors follow the 2MASS luminosity function more closely.

The role of AllWISE seems sub-dominant in this regime, compared to that of the upper limit on VISTA magnitude errors. If the constraints on magnitude errors were loosened, then the depth of AllWISE would be the main limitation, but the VISTA pipeline magnitudes would still be unreliable. Therefore, fainter samples from the VHS and VIKING would need model-based forced photometry from auxiliary catalogues. 

\section{Public Optical Surveys and optical Xmatches}
\label{sec:optical}%
%Many are the wide-sky optical surveys publicly available on the web. 
In order to maximize wavelength coverage and reach a wide mapping of the Southern Sky, we selected three among the most successful wide-field, multi-band optical photometric surveys: the Dark Energy Survey \citep[DES, ][]{Abbott18}, the Panoramic Survey Telescope \& Rapid Response System 1 (PanSTARRS1, hereafter PS1, \citealt{Chambers16}) and the SkyMapper Southern Sky Survey \citep{Wolf18}.  

For this work, we retrieved the latest public releases (DES-DR1, PS1-DR2, SkyMapper-DR1) and cross-matched them onto the final VISTA table of Section~2.
All matches have been performed with \textsc{TOPCAT}, using the 'Join - Pair Match'. 
Unless differently specified, we used the 'Sky' Matching algorithm with a maximum tolerance of $3\arcsec$. 

\begin{figure}
\includegraphics[width=\columnwidth]{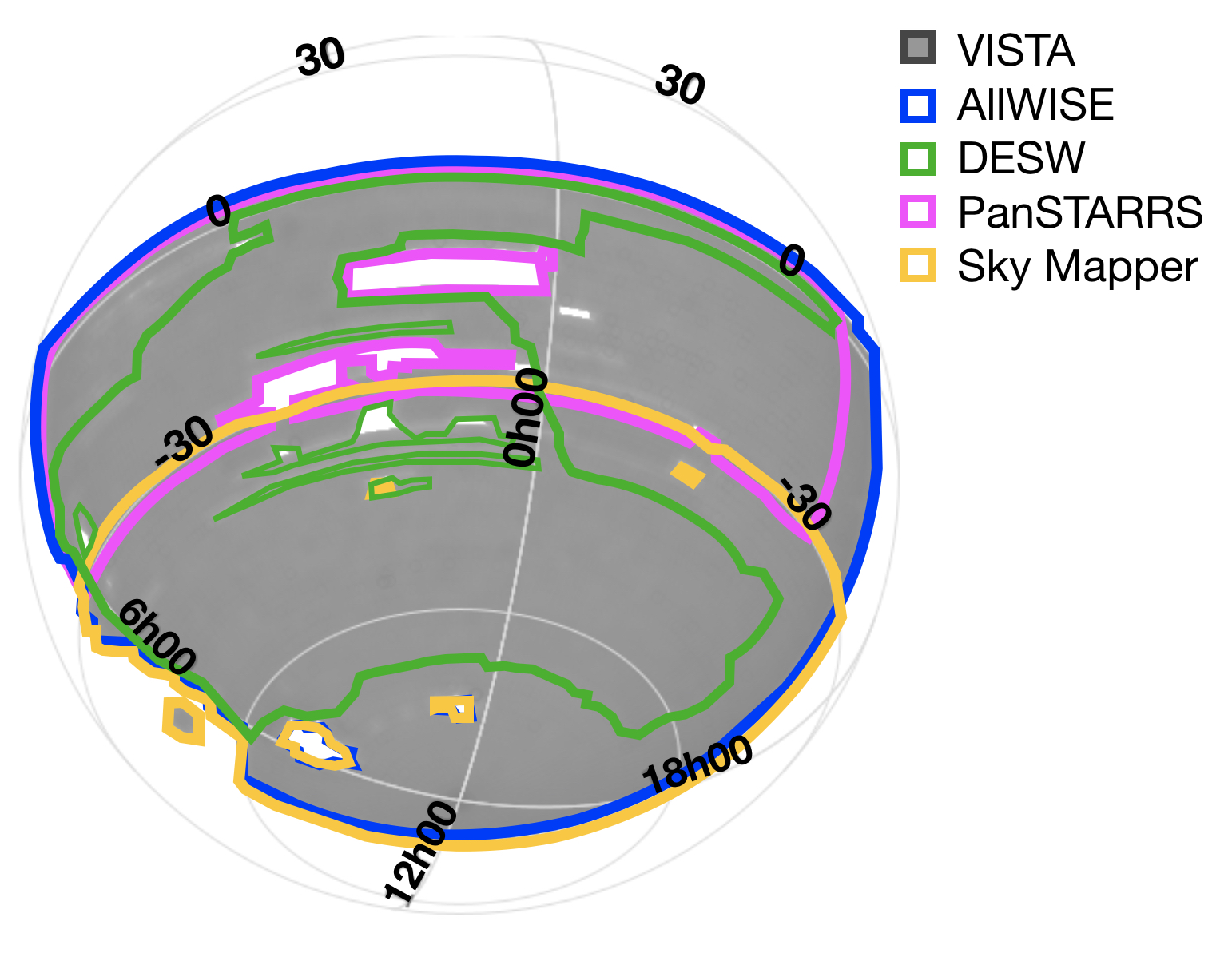}
\caption{Sky view of the VEXAS+optical table footprints, plotted over the VISTA catalogue footprint (grey).}
\label{fig:xmatches_coverage}
\end{figure}

The tables are further described in the following subsections. Their sky footprint is shown in Figure~\ref{fig:xmatches_coverage}, also including the coverage of the VISTA and AllWISE catalogues. The tables are publicly released and available for the scientific community through the ESO Phase-3 interface\footnote{for the moment being, while making the tables Phase-3 compliant, we release them in the common FITS format via a temporary repository (\href{https://drive.google.com/drive/folders/18IjYlkKrvEB2AEcj6UZiHTOUKbijULpf?usp=sharing}{here}).}. 

\subsection{VEXAS-DESW}
The DES covered 5000 deg$^2$ of the southern sky using the 570-Mega-pixel Dark Energy Camera \citep[DECam,][]{decam12}, mounted on the Blanco 4-meter telescope at Cerro Tololo Inter-American Observatory.
It provides photometry in $grizY$ filters, reaching limiting magnitudes (10$\sigma$: $g_{\rm lim}=24.45$, $r_{\rm lim}=24.30$, $i_{\rm lim}=23.50$, $z_{\rm lim}=22.90$, $Y_{\rm lim}=21.70$) faint enough to detect about 100 million stars, 300 million galaxies, $\approx10^5$ galaxy clusters as well as some $z>6$ quasars \citep{des16}. 

The VEXAS-DESW final table comprises 37'615'619 objects with $grizY$ photometry from DES, near-IR from VISTA and mid-IR from WISE.  
In particular, all the objects have at least one detection among the 
$grizY$ bands, one among the VEXAS $YJHKs$ bands and they all have a match in AllWISE, {with at least one measured WISE magnitude.  
Finally, 1'971'448 objects have reliable photometry (within the $10\sigma$ limiting magnitudes) in all optical and infrared bands plus $W1$ and $W2$} \footnote{The bottle-neck is the Y band from VISTA which is measured for only 2'402'713 sources}. 

For each entry and for each survey (VISTA and DES), we provide the right ascension and declination\footnote{Coordinates are given in the original format: radiants for VISTA and degree for all the other surveys.}, the source ID, optical and infrared magnitudes with corresponding errors ($grizy$ from DES, $YJHKs$ from VISTA). We also give, for each entry in the table, the separation, computed by \textsc{TOPCAT}, between the centroid in DES and the centroid in VISTA.  In addition, we also provide the $W1$ and $W2$ magnitudes from AllWISE, the galactic dust extinction value measured from the Schlegel maps \citep{Schlegel98}, plus a number of columns that quantify the extendedness of the source. 
In particular: \texttt{PSTAR}, the probability that the source is a star from VISTA imaging; and \texttt{spread\_model\_i}, \texttt{spreaderr\_model\_i}, and \texttt{wavg\_mag\_psf\_i} from DES. 
The \texttt{spread} indicator values and errors have been used by the DES collaboration to identify galaxies \citep[as detailed by][]{Vikram15}. Alternatively, the difference \texttt{wavg\_mag\_psf-wavg\_mag\_auto}, also provided, between PSF-like and isophotal magnitudes, can be used to separate point-like and extended sources.

\begin{figure}
\includegraphics[width=8cm]{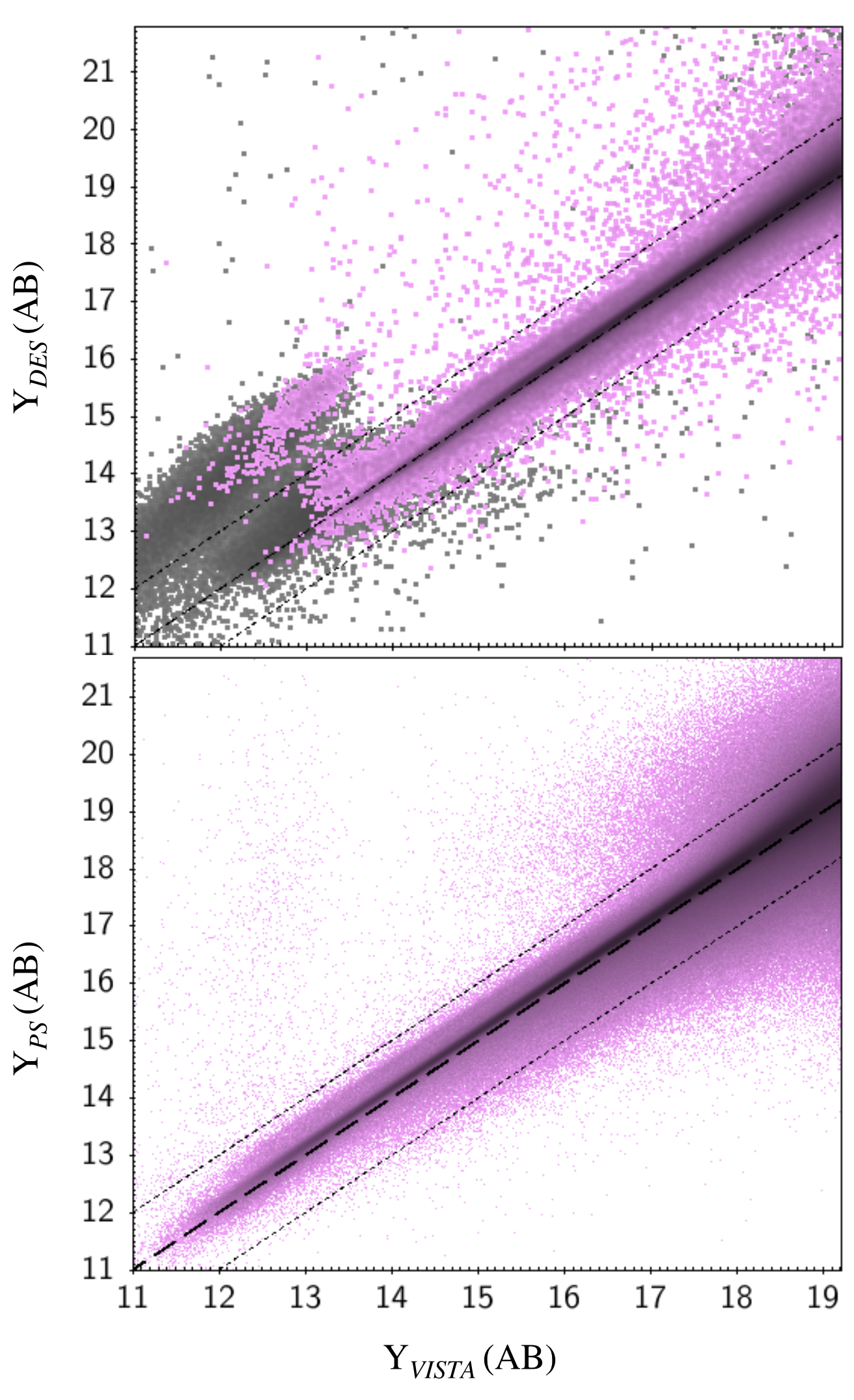}
\caption{Direct comparison between $Y-$band magnitudes from optical and infrared surveys, in the AB reference system, for objects with reliable photometry in an overlap of VISTA and DES/PanSTARRS. The thick dashed lines follow $Y_{PS/DES}=Y_{VISTA},$ and the dotted lines delimit offsets of one magnitude,%. A fair agreement ($-1<Y_{VISTA}-Y_{DES}<1$) is found for 99\% of the sources.
enclosing $>95\%$ of the sources. A small systematic difference can also be attributed to the slightly different filters used in the different surveys.  The comparison between DES and VISTA Y-band magnitude (upper panel) shows $\approx$15k bright, saturated objects with $Y_{DES}>Y_{VISTA}+1.$ 
This is less apparent in PS1, due to its smaller primary mirror and hence fewer saturation issues. The grey-coloured (resp. magenta) objects have $Y_{VISTA}$ magnitude errors larger (resp. smaller) than $0.5$~mag.}
\label{fig:yDESvsyVISTA}
\end{figure}

We caution the reader that the magnitudes in the tables are provided in their native system of reference (AB for DES, Vega for VISTA and WISE). However, in Figure~\ref{fig:yDESvsyVISTA} (upper panel), in order to provide a direct comparison between the $Y$-band magnitudes measured from DES and those measured for the same band from VISTA, we translate the latter in the AB system, using the following simple equation: $Y_{AB} = Y_{Vega}+0.643$\footnote{We used standard Vega-to-AB conversions, summarised here: \url{http://www.astronomy.ohio-state.edu/~martini/usefuldata.html}}. 
A fair agreement can be found between for 99\% of the $\approx 1.4$M objects for which the $Y$-band is measured in both surveys. 
Nevertheless few differences can be seen at the brightest and faintest end. 
In particular, the VISTA magnitudes appear to be systematically brighter for $\approx18$k objects ($Y_{VISTA}-Y_{DES}<-1$), of which $\approx15$k are bright stars ($Y_{VISTA}<13$, PSTAR$>0.85$) with however a large error on the $Y_{VISTA}$ magnitude ($>0.5$ mag). In the figure we plot in purple the objects with  $\delta Y_{VISTA}<0.5$ and $\delta Y_{VISTA}<0.5$, and in grey objects with larger error values in one or both surveys. It is clear that considering only objects with small magnitude errors alleviates this disagreement but does not resolve it.  
As a direct, visual inspection of $1000$ randomly selected objects confirmed, they are mostly saturated sources. We can then ascribe the $Y-$band mismatch to artificially faint magnitudes estimated by the DES pipeline on saturated objects. 

\subsection{VEXAS-PS1W}

The Pan-STARRS1 (PS1) comprises the first two data releases of images taken with the PS1 1.8-meter telescope at the Haleakala Observatories in Hawaii, equipped with a 1.4 Gigapixel camera \citet{Kaiser02}, which covered the entire sky north of $Dec=-30$ deg. Its $grizy$ filter set is comparable to that of the DES, although it is somewhat shallower (5$\sigma$: $g_{\rm lim}=23.3$, $r_{\rm lim}=23.2$, $i_{\rm lim}=23.1$, $z_{\rm lim}=22.3$, $Y_{\rm lim}=21.4$).

The VEXAS-PS1W final table comprises %29'873'392 
24'693'386 objects with optical photometry from PS1 and infrared photometry from VISTA.  
As in the previous case, all the objects have at least one magnitude entry among the $grizY$ bands and at least one among the VEXAS bands $YJHKs$.  
Of these, 2'328'832 %5'463'435 
have photometry (within the 5$\sigma$ limiting magnitudes) measured in all nine bands. 

Also for VEXAS-PS1W we provide source ID, coordinates and magnitudes (with associated errors and in the original reference system) for both surveys, the separation between the centroids, plus the probability to be a star and the galactic dust extinction value from VISTA. For PanSTARRS DR1 we give Petrosian magnitudes in each of the optical bands and also the PSF $i$-band magnitude, its associated errors and the stellarity likelihood.  Moreover, as already done for VEXAS-DESW, all the entries have also a match in AllWISE and thus we provide the $W1$ and $W2$ magnitudes in the Vega magnitude system.  

The lower panel of Figure~\ref{fig:yDESvsyVISTA} shows the AB $Y$-band comparison between PanSTARRS and VISTA. Also in this case, the majority (96\%) of the objects have $-1<Y_{VISTA}-Y_{PA}<1,$ among sources with a trustworthy $Y$-band magnitude value ($\approx$ 8M object, within the survey nominal limiting magnitude). 
The scatter becomes larger at magnitudes fainter than $Y>17$ but a fair agreement is found between the two measurements. We note a small systematic difference with $Y_{PA}$ of $\sim0.2$ mag fainter than $Y_{VISTA}$. This is mostly driven by the different shape of the $y_P$ and $Y$ filters used in PS1 \citep{Tonry2012} and VISTA.

Due to its smaller primary mirror ($\approx1.8$m versus $\approx4$m), PS1 is less affected than DES by the saturation issues described in the previous section and shown in the upper panel of the figure.

\subsection{VEXAS-SMW}

The Australian SkyMapper Southern Sky Survey\footnote{Led by the Research School of Astronomy and Astrophysics at the Australian National University, in collaboration with seven Australian universities and the Australian Astronomical Observatory.} aims at producing a homogeneous multi-band coverage of the Southern sky, similar to the one that the Sloan Digital Sky Survey provided in the Northern sky. 
The first data release covers an area of $\approx17000$ deg$^2$ in six pass-bands ($uvgriz$) with point-source completeness limits in AB magnitudes of 17.75, 17.5, 18, 18, 17.75, 17.5 for $u,$ $v,$ $g,$ $r,$ $i,$ $z,$ respectively. 

The Xmatch between VISTA and Sky Mapper produces 20'331'041 objects, 
but however only 82'052 objects have measured photometry in all nine bands ($ugrizYJHKs$). The $H$-band from VISTA and the $u$-band from SM are the two least covered bands with $\approx 3.2$ and $\approx 3.8$ of objects respectively. 

In the VEXAS-SMW final table we provide, again, coordinates and magnitudes for the two surveys, and the WISE ($W1,$ $W2$) magnitudes. Two values for the galactic dust extinction, calculated with the same maps \citep{Schlegel98}, one for each survey, are provided. The difference between the two values is that the VISTA one uses the correction given in \citet{Bonifacio00} that reduces the extinction value in regions of high extinction (i.e. $E(B-V)>0.1$).
Finally, besides the \texttt{PSTAR} stellarity from VISTA, we also provide the value \texttt{class\_star} for Sky Matter. In principle, both values should quantify the likelihood of an object to be an unresolved source. However they depend on the model that has been fitted to the data (in a given detection band) and also on the depth and image quality of the survey. A comparison between these two values for bright objects in  $i$-band (46'832 sources with $i<15$ in the AB system, i.e. the peak of the $i-$band magnitude histogram), shows a fair agreement between \texttt{PSTAR} and \texttt{class\_star} for 46'046 high confidence stars. Among the remaining objects, 519 are classified as secure stars in SM (\texttt{class\_star}>0.99) but have very low \texttt{PSTAR} and 8 objects have instead \texttt{PSTAR}>0.95 but \texttt{class\_star}<0.4. Finally, 150 sources have \texttt{PSTAR}<0.075 but \texttt{class\_star}>0.3. Using both stellarity indicators may help classifying objects with extended morphology and appreciable colour gradients, e.g. quasars with bright host galaxies.

\subsection{VEXAS-EXTRAGAL}
\label{sec:extragalactic}
As an ancillary by-product, we also produced a table of 5'742'300 objects with extragalactic colours in the VISTA footprint. We do so by requiring
\begin{equation}
    W1-W2>0.2+\sqrt{\delta W1^{2}+\delta W2^{2}}
\end{equation}
on the WISE magnitudes (in the Vega system). This colour cut excludes most stellar objects, which have $W1-W2\approx 0.$ It generalizes the exponential cut-offs proposed by \citet{Assef13} to isolate AGN candidates, which had a more demanding threshold at $W1-W2>0.7$ and were incomplete with respect to quasars at redshifts $z\gtrsim2.$ Our colour-cut also retains most galaxies at $z\lesssim2$ (which have $W1-W2\approx0.35$).
We note that we consider only the first two WISE bands, since the sensitivity in $W3$ and $W4$ is significantly impacted by Earth-shine noise.

A fraction of contaminants include brown dwarfs and white dwarfs with an IR excess, which can be further skimmed if optical and NIR magnitudes from ground-based surveys are used. In general, more refined separations of galaxies, stars and quasars can be performed if multi-band coverage is available (e.g. through optical-to-NIR colours). However, this is currently possible only for a part of the Southern Hemisphere. 

For the VEXAS-EXTRAGAL table, we provide coordinates and magnitudes for the VISTA and AllWISE surveys, the match and the magnitudes for 2MASS, when available, as well as stellarity index (PSTAR) and the galactic dust extinction values.

\section{Spectroscopic optical Xmatches}
\label{sec:spec_xmatches}
In multi-band object classification and discovery, the lack of spectroscopic information is a common issue. This holds especially in the Southern Hemisphere, where (at present) spectroscopic surveys are affected by depth, footprint and preselection. However, part of the existing spectroscopic surveys do overlap with our VEXAS footprint.

For this reason, %as a further service to the scientific community, 
we performed a spectroscopic Xmatch between the final VISTA table and two of the most used all-sky spectroscopic surveys: the  Sloan Digital Sky Survey (SDSS, DR14\citealt{SDSS_DR14}) and the The 6dF Galaxy Survey (6dFGS, \citealt{6dfgs_I}, DR3 \citealt{6dfgs_DR3}).

From both surveys, we only retrieved objects with a secure spectroscopic redshift measurement, using the 'quality' keyword for 6dFGS (i.e. only objects with \texttt{qual=3} and \texttt{qual=4}, corresponding to 110'256/136'304 objects) and the error on the SDSS redshift value ($z_{\rm err}<0.01$, 4'716'522 objects). 

\begin{figure}
\includegraphics[width=8cm]{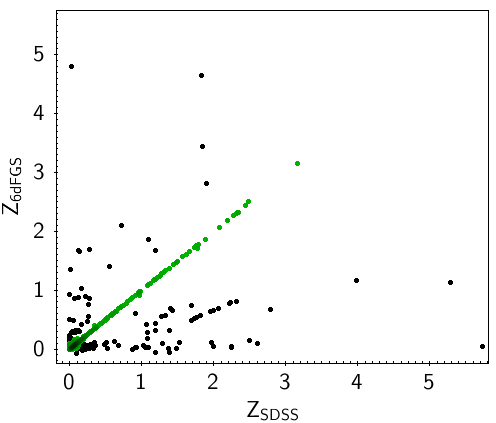}
\caption{Comparison of the redshifts computed from 6dfGS and these computed from SDSS for the common sample of $2834$ objects. A good agreement is found for the 2709 green points that lie on the one-to-one line, however for $\sim5$\% of the sample  (black points), the two survey report very different redshift estimates for the same object .}
\label{fig:redshift_comparison}
\end{figure}

The Xmatch produced 56'768 entries for 6dFGS and %92'223 
347'076 for SDSS, of which 2834 %519 
objects have a double common match. These objects, with one or more spectroscopic classifications are contained in the table 'VEXAS-SPEC.fits', which is released as ancillary product of the VEXAS DR1. 
The 2855 systems with a double match allow us to directly compare the redshift estimates that the two surveys provide on the same objects, as displayed in Figure~\ref{fig:redshift_comparison}. For $\sim95$\% of the sample with a double spectroscopic match, the two surveys agree well on the measured redshifts (green points, 2709/2834). However for the remaining $\sim5$\% of the matches, SDSS and 6dFGS strongly disagree on the redshift inference. 
The disagreement can be due to multiple reasons: blends of multiple objects in the same (slightly offset) fibre diameter, over-correction of telluric absorption, splicing of the red and blue fibre spectra, and catastrophic failures in the pipelines for emission/absorption-line redshifts. 
This problems leads e.g. to degenerate solutions for quasars at $z_s\approx0.77$ and $z_s\approx2.2,$ in the low signal-to-noise regime. 

\section{Other wavelength Xmatches}
\label{sec:otherwave}
To further extend the wavelength coverage of VEXAS, we also performed a Xmatch with surveys in the radio and X-Ray domain. 
Also in this case, 
we require the existence of match in the AllWISE catalogue (within $10\arcsec$) for each object, in order to eliminate spurious detections. 
Thus, for the X-Rays, we retrieve directly the catalogues presented by \citet{Salvato18}, who performed a maximum-likelihood match of sources from the ROSAT all-sky survey (RASS, \citealt{Rosat_I, Rosat_II}) and the XMM-Newton with the AllWISE Source Catalogue.% described in Sec.\ref{sec:WISE}.
In the radio domain, instead, we retrieved 21cm detection data from the Sydney University Molonglo Sky Survey (SUMSS, \citealt{Sumss_I}), and performed the matches ourselves.

{We emphasize that both for radio and X-ray we keep the WISE matching radius to $10\arcsec$, rather than restricting it to $3\arcsec$ (as done in the optical), given the poorer resolution of the matched surveys. This also preserves the maximum-likelihood matches performed by \citet{Salvato18} between WISE and X-ray detections.}

After cross matching with VISTA, we obtained $\approx 37$k objects with a match in another wavelength. Of these, 139 objects have both a match in X-ray and in 21cm catalogues, 3049 have a match in X-Ray (1937 from 2RXS and 1147 from XMM) and 34192 have a 21cm match. 
In the following we describe the Xmatch procedures at larger and smaller frequencies providing a brief description of the resulting tables, which are publicly available from VEXAS. 

\subsection{VEXAS-XRAYs}
We cross-matched VISTA with two among the most successful X-Ray all-sky surveys in the 0.1 - 2.4 keV band: the ROSAT and the XMM-Newton Surveys. 

The ROSAT all-sky survey (RASS) has been the first observation of the whole sky in the 0.1 - 2.4 keV band \citep{Truemper82}. 
The RASS bright source catalogue \citep{Voges99}, containing 18806 sources, and the RASS faint source catalogue \citep{Voges00} containing 105'924 sources down to a detection likelihood limit of 6.5 were joined together in the ROSAT 1RXS catalogue \citep{Rosat_I}, which was then finally reprocessed in \citet{Rosat_II} (2RXS) with a new improved detection algorithm that allowed to significantly reduce the number of spurious sources.
The total number of entries listed in the 2RXS catalogue is 135'118. 
Of these, 5926 have been flagged as uncertain detections and have been therefore excluded from our catalogue. The match of the 2RXS catalogue with AllWISE (within $10\arcsec$) produced 127'361 matches, that we then Xmatched with the VISTA final table finding 1937 2RXS detections. 

The cross-match between XMM-slew and AllWISE produces 17'834 matches, of which only 1147 have also a match in VISTA. 
The objects in common between 2RXS and XMM are 40, thus yielding 3044 unique sources with an X-Ray. These sources are reported in the publicly released 'VEXAS-XRAYs' table. For each entry in the table we 
%The table has 3044 (which 40 detected in both surveys) entries for which we
report: coordinates for VISTA (in radiants), for AllWISE and for the parent X-ray survey (in degree), infrared magnitudes with corresponding errors and X-Ray fluxes, separation between the centroids of the different surveys, the galactic dust extinction values and the stellarity likelihood from VISTA. 

\subsection{VEXAS-21cm}
The Sydney University Molonglo Sky Survey (SUMSS, \citealt{Sumss_I, Mauch03}) is a radio imaging survey of the southern sky (dec<-30~deg), covering 8100~deg$^2$ at 843 MHz with images with resolution of $43\arcsec\times43\arcsec$ cosec(Dec.), rms noise level of about 1 mJy/beam and a limiting peak brightness between 6~mJy/beam and 10 mJy/beam.
The final catalogue is complete to 8 mJy at dec.$\le -50$~deg, and to 18 mJy at dec$> -50$~deg. 
Positions in the catalogue are accurate to within $2\arcsec$ for sources with peak brightness $\ge20$~mJy/beam, and are always better than $10\arcsec$. 
The internal flux density scale is accurate to within 3\%. 

We queried the SUMSS catalogue table directly\footnote{Available through the CDS portal, table \texttt{VIII/81B/sumss212}}, and cross-matched it locally with our VEXAS-AllWISE3 by seeking the closest source within a 10$^{\prime\prime}$ matching radius. This yielded 77'338 radio sources with a match in VISTA and AllWISE, which we report and release in the table "VEXAS-21cm". 

\section{Scientific Cases}
\label{sec:science}
In the following sections we describe few scientific applications using the VEXAS tables to search for peculiar and rare objects in the sky, such as strong gravitationally lensed quasars or extremely red objects. 

\subsection{Extremely red objects}
\label{sec:eros}
Extremely red objects (EROs) are massive high-redshift galaxies that have been identified in the infrared and whose colours are so red (R-K $> 5$ or I-K $>4,$ Vega system) that they are not found in surveys that select galaxies at visual wavelengths. 
\begin{figure}
\includegraphics[width=\columnwidth]{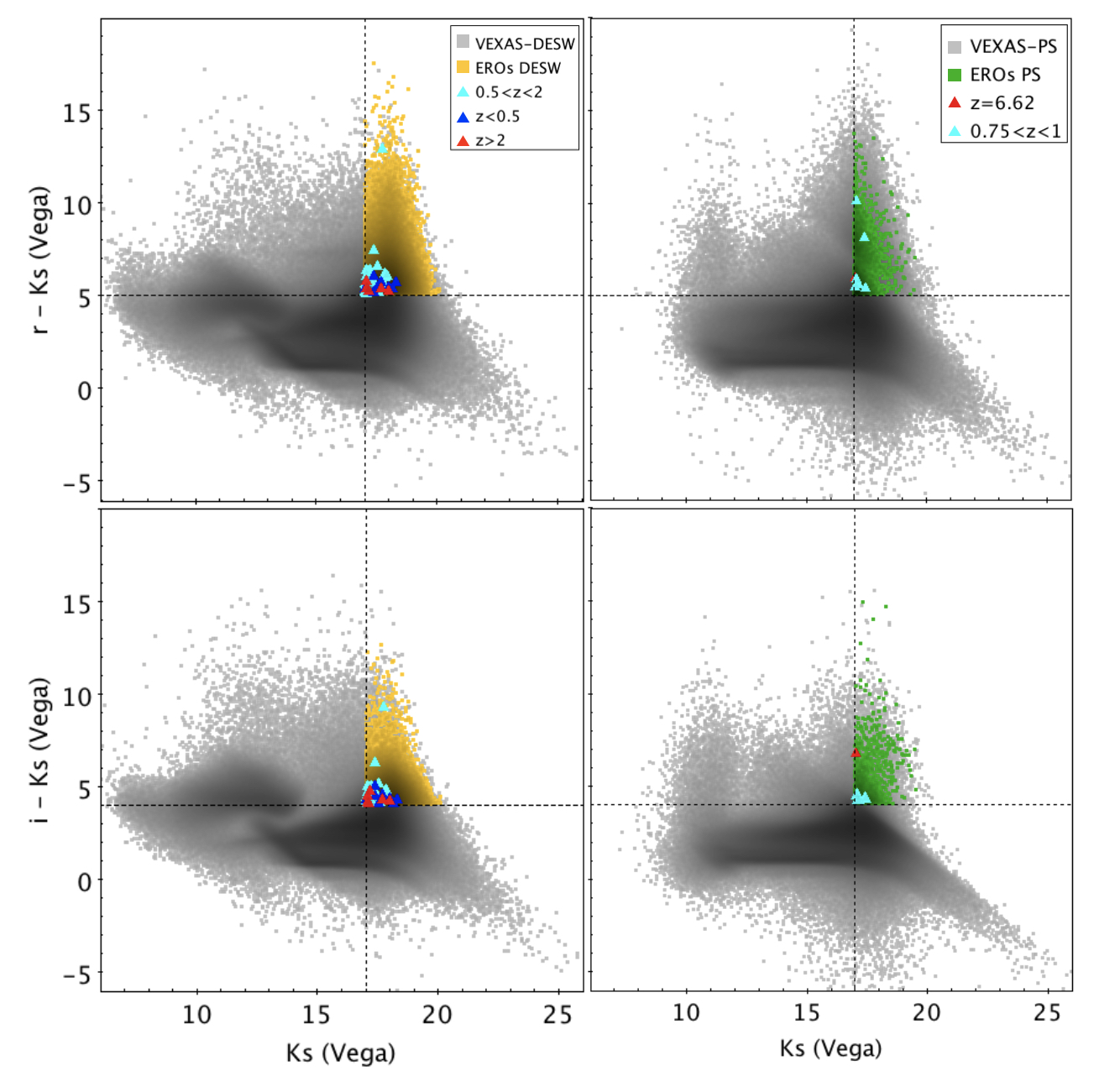}
\caption{EROs candidates selected from the VEXAS-DESW catalogue (left, yellow) and from the VEXAS-PS catalogue (right, green) using the magnitude-colours thresholds defined in the text, highlighted in the plots with dotted lines. Triangles are the spectroscopically confirmed objects from the VEXAS-Spectroscopy table (59 in total).}
\label{fig:EROs_selection}
\end{figure}

The first identification of EROs goes back to the first time 
that a deep survey of the $2\mu$m sky was carried on. 
\citet{Elston88} reported the discovery of a new population of extended objects that occupy a region of the infrared colour-magnitude diagram that is separate from the locus of normal galaxies. 
In particular, they identified a group of galaxies with K magnitudes of  $\approx17-18$ and R-K$\ge5.$ Those authors suggested that these objects may be high-redshift galaxies ($z>$6) undergoing a luminous star-forming phase, or weakly evolved ETGs at $z\sim1 $--$1.5$.
Since the late eighties, large effort has been devoted to surveying the infrared sky and many more EROs have been found \citep{McCarthy92, Thompson99, Daddi00}. 

The red colours of EROs are consistent with two classes of galaxies: (\textsc{i}) old passive galaxies at $z\sim1$ with a large K-correction; or (\textsc{ii}) high redshift star-forming galaxies or heavily dust-reddened AGN. 
Observations suggest indeed that both these two classes of objects populated the EROs class: few objects were spectroscopically confirmed to be dynamically relaxed and passively evolving early type galaxies at $z\ge1$, (pEROs, e.g. \citealt{Stiavelli99,Kong06}), but many others were detected in the sub-mm, thus identified as high-redshift starbusts reddened by strong dust extinction and characterized by high star formation rates (sfEROs, e.g. \citealt{Andreani00}). 
pEROs turned out to be the most massive (M$\ge 10^{11.5}$M$_{\odot}$; \citealt{Saracco05}) and old (age $>1$ Gyr; \citealt{Cassata08}) galaxies at $z > 1.$ In addition, the pEROs show strong clustering that is comparable to that of local massive ellipticals (e.g., \citealt{Daddi00, Roche02, Brown05, Shin17}). 
This strongly suggests that pEROs could be the progenitors of local massive ellipticals and that they can work as powerful tool to study the redshift evolution of the most massive galaxies in the local universe and thus to test our current cosmological galaxy formation models (e.g., \citealt{Gonzalez11}).
The relative contribution of the two classes of objects to the ERO population is still unknown, thus it is of great importance to build a statistically large sample of candidates to enlarge the ERO population and constrain number counts and density, especially for pEROs, with high precision. 

Thanks to our Xmatched tables, one can now build a catalogue of EROs candidates and also put some constraints on the relative number of pEROs and sfEROs, performing a match with the VEXAS spectroscopic table. 
Following previous studies in the literature, we selected objects with:
\begin{equation}
Ks \ge 17 \ ,\ Ks - i > 4 \ ,\ Ks - r > 5\ .
\label{eq:eq_eros}
\end{equation}

We applied these selection criteria to both the VEXAS-DESW and the VEXAS-PS final tables, finding a total of 1'010'205 EROs candidates (8758 from PS, 1'001'498 from DESW and 51 in common)\footnote{The table 'EROs-candidates.fits' is released as ancillary product of the VEXAS DR1.}. 
Then, Xmatching this table with the VEXAS spectroscopic table, we also found 59 objects with a spectroscopic redshift and a secure classification. Among these, 26 are classified as QSOs and 7 have redshift $z>2$ while the remaining 33 are classified as GALAXY and have $0.19<z<1.64$. 

Figure~\ref{fig:EROs_selection} shows the candidates from the two surveys in the magnitude-colour plots used for the selection. The spectroscopic objects are highlighted as triangles, colour-coded by their redshifts, as explained in the caption.

\subsection{Astrometric offsets and lensed quasars}
\label{sec:lenses}
The astrometric consistency of sources across different catalogues can be used to identify close blends of multiple objects. One application is the detection of gravitational lenses, which are often blended in ground-based catalogue surveys due to their typical image separations and survey image quality.

In particular, strongly lensed quasars (QSOs) are rare objects that require a close alignment between a quasar and a massive galaxy. According to the estimations of \cite{Oguri10}, this happens once in $\sim10^4-10^3$ times \citep[see][for revised estimates]{AS19}. 
The most successful ways to look for lensed QSOs are based on multi-colour pre-selection of QSOs and galaxies from wide-field surveys, followed by morphological criteria to isolate the rare cases of veritable lenses. 
For the first step, as largely demonstrated from previous work (and recalled in Sect.~\ref{sect:known_lenses} below), a wide wavelength baseline is crucial to properly disentangle high-z quasars from stars and galaxies (e.g. from blue, optical bands to the mid-infrared). Thus, a homogeneous multi-wavelength coverage on the whole SGH like VEXAS is ideal for this purpose. 

For the second step, we use the very basic assumption that if the deflector and quasar images contribute differently in different bands, this should result in a centroid offsets of the same object among different surveys. To this aim, we developed in a previous work \citep{Spiniello18,AS19} the \textit{BaROQuES} scripts\footnote{Blue and Red Offsets of Quasars and Extragalactic Sources, available upon request, \texttt{https://github.com/aagnello}} and we apply them here to the VEXAS-ESM and the VEXAS-EDESW where, thanks to the WISE colour-cuts, we have selected extragalactic objects, thereby eliminating the most stellar contaminants.The BaROQuES consider \textit{field-corrected} offsets: for each object, the astrometric offsets between different survey catalogues are computed in patches of $0.5\times0.5$~deg$^{2}$ around it, and their average is subtracted to the offset on the central object. This is done in order to remove any systematics from atmospheric differential refraction, across surveys with different waveband coverage.

\begin{figure*}
\includegraphics[width=18.5cm]{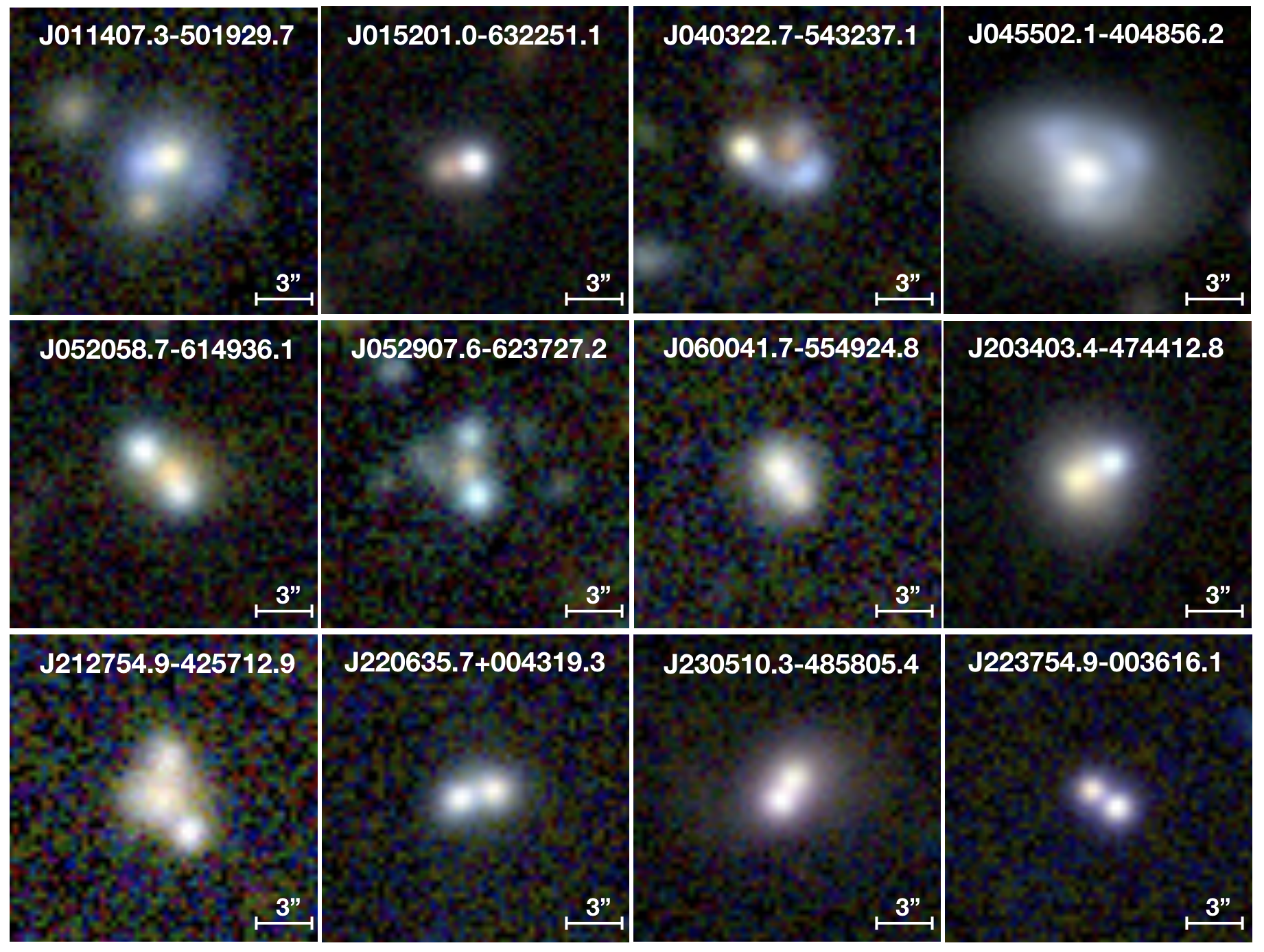}
\caption{Examples of gravitational lens candidates found through field-corrected offsets between their DES and VISTA coordinates.}
\label{fig:cands}
\end{figure*}

When DES (resp. SkyMapper) is combined with VISTA, $\approx5\%$ of the objects have field-corrected offsets above $0.26\arcsec$ (resp. $0.4\arcsec$). The threshold is comparable to the optical survey pixel size, but it also depends on the survey depth, which contributes to the astrometric accuracy on ground-based astrometry. In particular, for the DES, the sample can be further reduced (from $\approx5\%$ to $\approx3\%$) if a magnitude-dependent cut is used for objects fainter than $i\approx21.$ Approximately half of the known lenses in the VEXAS-DES footprint have field-corrected offsets above threshold, the other half being larger-separation systems with bright ($i\lesssim19$) catalogue magnitudes.  Figure~\ref{fig:cands} displays some of the previously unknown, gravitational lens candidates identified this way.
Magnitude-dependent offset cuts are (currently) not relevant for SkyMapper, due to its rather larger pixel size and shallower limiting magnitudes.

\begin{figure*}
\includegraphics[width=18cm]{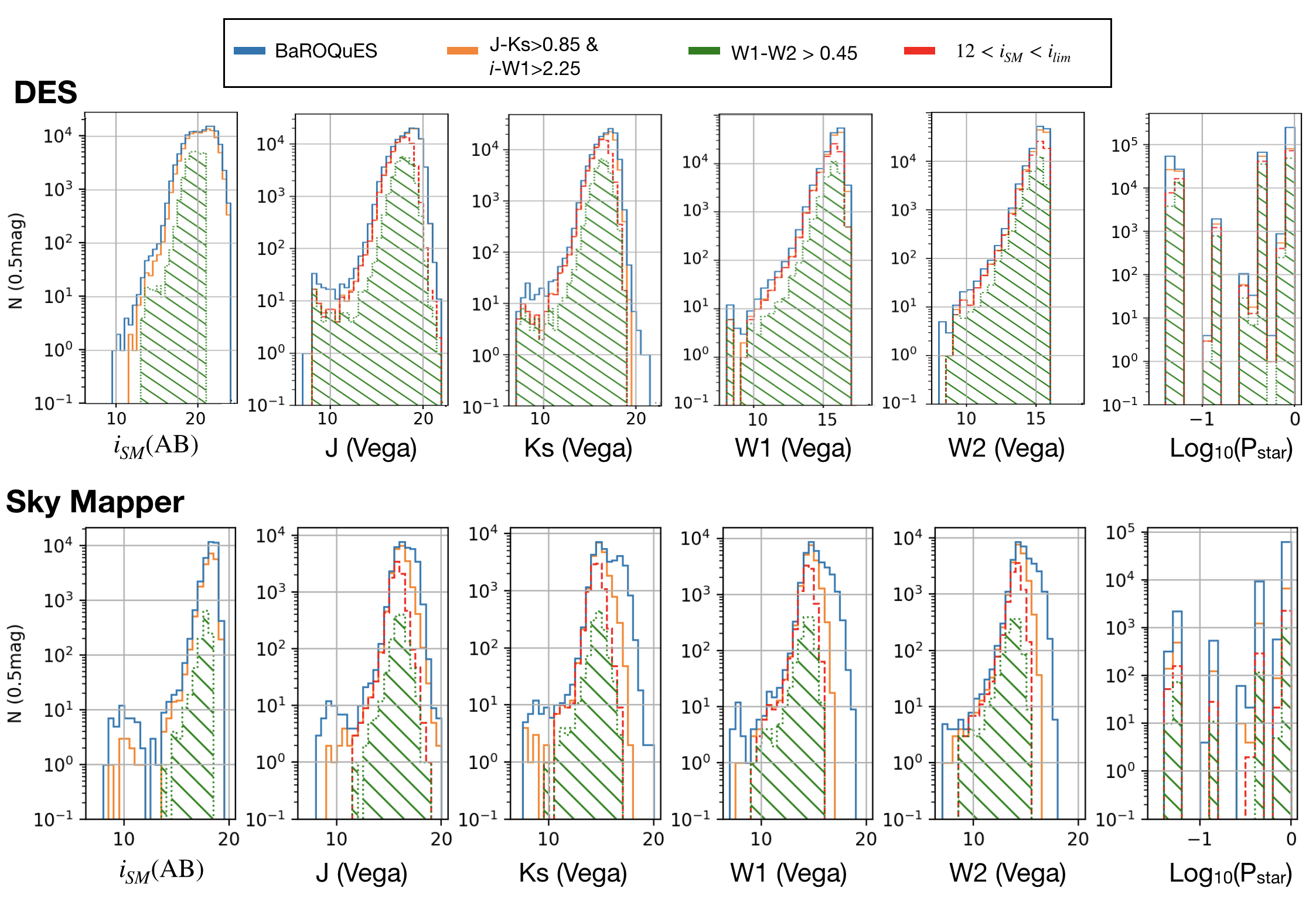}
\caption{Histograms of magnitudes and stellarity of objects with anomalous astrometric offsets in DES (\textit{top} row), and in SkyMapper with dec<-70 (\textit{bottom} row). Different nested histograms correspond to incremental cuts in hybrid colours, limiting magnitude, and strict extragalactic cut in WISE. The limiting $i-$band magnitude is chosen as $i_{lim}$=21 for DES, corresponding to current limits for spectroscopic follow-up on $\approx4\ m-$class telescopes, and $i_{lim}$=19 for SkyMapper, simply as the limiting depth of its DR1. WISE preselection is not a limiting factor for wide and shallow searches ($i\lesssim19$), but it does become the main limitation in deeper searches ($i\lesssim21$).
}
\label{fig:Baroques_histo}
\end{figure*}

Since different surveys are used to find close blends, the final sample selection can be affected by the limitations of each contributing survey. Figure~\ref{fig:Baroques_histo} shows histograms of $i-$band, $JHKs$ and WISE magnitudes (and a stellarity indicator) for a sub-sample of BaROQuES in the DES and SkyMapper footprints, through different cuts. The samples are mostly unaffected by simple extragalactic colour cuts ($J-Ks>0.85,$ $i-W1>2.25$) or by a WISE colour cut. The main limiting aspect is the depth of the parent surveys: the depth of SkyMapper dominates the preselection, whereas for DES the main limitation is from the depth of WISE.

\subsubsection{Known lenses in VEXAS}
\label{sect:known_lenses}
Thanks to VEXAS, we can also better characterize the population of already known lenses to facilitate future searches. 
We assembled a table with 277 lenses collected by the CASTLES Project \citep{Munoz98} database, the SDSS Quasar Lens Search (SQLS, \citealt{Inada12}) or recently discovered by us \citep{Agnello15, Agnello18_atlas, Agnello18_gaia, Spiniello19_des,Spiniello19b} and other groups (e.g. \citealt{Ostrovski18, Anguita18, Lemon18}). % matching, indeed, wide-sky surveys. 
Among these, 69 are in the VEXAS footprint, 48 are present in VEXAS-DESW, 26 in VEXAS-PS, 25 in VEXAS-SM and finally 65 also have a match in AllWISE. 

\begin{figure*}
\includegraphics[width=18.5cm]{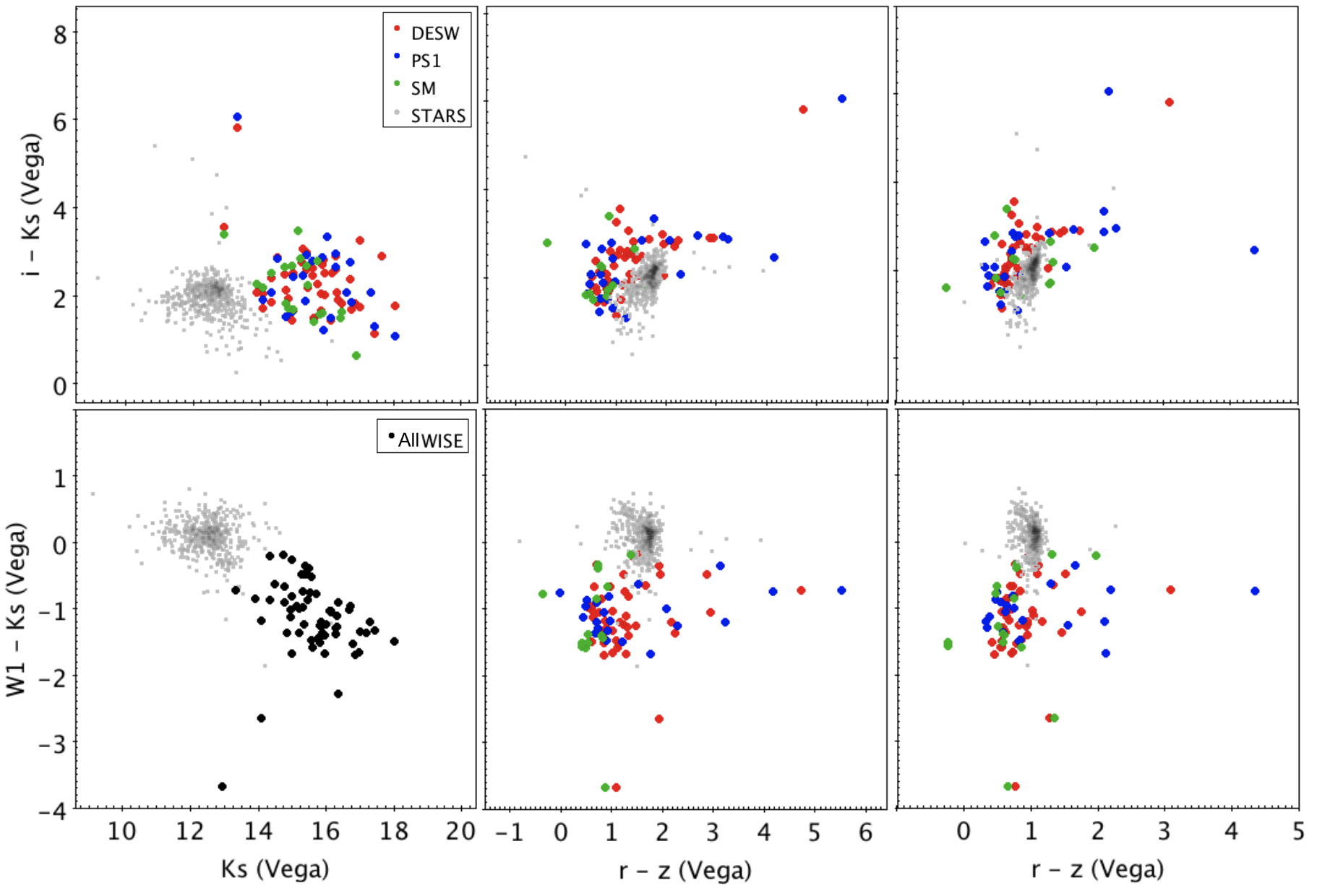}
\caption{Known lenses from the literature in colour-magnitude and colour-colour multi-wavelength plots. Different colour coding identifies lenses recovered in different VEXAS sub-tables, as explained in the captions. Grey small points are objects with a spectroscopic match and class=STARS }
\label{fig:Known_lenses}
\end{figure*}

Figure~\ref{fig:Known_lenses} shows colour-magnitude or colour-colour plots of these known lenses in VEXAS. 
The known lenses typically span regions outside the stellar loci, and are generally characterised as extragalactic objects. While most have quasar-like hybrid colours, this is also a by-product of preselection in the (older) campaigns that identified them.

\subsection{Galactic and Solar System science}
\label{sec:galactic-science}
Most of the discussion above has been centred on extragalactic applications. However, exploitation of our VEXAS catalogues is not limited to extragalactic science. For example, one aim of the NeoWISE mission was to identify near-Earth objects and asteroids \citep{Mainzer11}. High proper-motion white dwarfs were among the first results by PanSTARRS1 \citep{Tonry2012b}. The disc and halo of the Milky Way have been mapped with dwarf and giant stars through 2MASS and optical data, also at regions of high galactic latitude \citep[e.g.][]{Maj03,Boc07,Gil12,Car19}. Still in the South, various faint streams and Milky Way satellites have been discovered in the DES footprint \citep[see e.g.][]{Drl15,Shipp18}.

As shown also by our Table~\ref{tab:numb_objects}, out of $\approx138$~million objects in the VISTA+AllWISE footprint, only  $\approx6$~million have extragalactic WISE colours. The increased depth and coverage of VEXAS with respect to 2MASS and previous releases of the VHS should therefore allow for extensions of Galactic and Solar-system studies towards fainter magnitudes and with extended multi-band coverage.

\section{Conclusions and future releases}
\label{sec:conclusions}
In this paper we have presented the first data release of VEXAS: the VISTA EXtension to Auxiliary Surveys (VEXAS DR1). 
We built the largest and most homogeneous multi-wavelength cross-matched photometric catalogue covering the Southern Hemisphere and containing only high-confidence objects with a match in at least two different surveys. 
In particular, we have: 
\begin{itemize}
\item used infrared multi-band photometry from the two wide sky main VISTA Surveys, VHS \citep{McMahon12} and VIKING \citep{Sutherland12}, building the VEXAS input table, containing 198'608'360 objects having good photometry in at least one VISTA band; 
\item required a close match with the AllWISE Source catalogue \citep{Cutri13}, to avoid as much as possible spurious detections and also cover the mid-infrared bands at 3.4$\mu\mathrm{m}$ and 4.6$\mu\mathrm{m}$;
\item cross-matched the VEXAS final table with three among the most used optical wide-sky photometric surveys (the DES, PanSTARRS1, SkyMapper), two X-Rays surveys (the ROSAT All Sky Survey, and the XMM-Newton Serendipitous Survey) and a 21cm survey (SUMSS, \citealt{Sumss_I}). 
\item cross-matched the VISTA final table with two state-of-the-art spectroscopic surveys, the Sloan Digital Sky Survey (SDSS, DR14, \citealt{SDSS_DR14}) and the the 6dF Galaxy Survey (6dFGS, \citealt{6dfgs_I}), to provide test sets for object classification and photometric redshift estimation.

\item showed the scientific potential of the VEXAS X-matched tables, presenting two extragalactic science cases to search for rare objects in the sky (namely strong gravitational lenses and extremely red objects). 
\end{itemize}

The extension of near-IR catalogue tables (from VISTA) to magnitudes in other wavebands enhances the purity and completeness of object classification, ensuring wide and homogeneous investigation into galaxies, stars, quasars and peculiar objects. The depth of VISTA naturally extends the scientific scope of research that relied on shallower magnitudes from e.g. 2MASS, and the cross-match with optical surveys enables a first characterization of objects with extreme SEDs.

\subsection{Future developments}
This first release included only catalogue cross-matches and was restricted to the Southern Galactic Hemisphere ($b<-20$~deg). The only space-based mission included here is WISE, due to its all-sky coverage and fair depth. This release does not include regions that strongly overlap with the Galactic disc ($-20\mathrm{deg}<b<200\mathrm{deg}$), which were already processed to a more advanced stage through the VVV survey \citep{Minniti10}.

While the current VEXAS catalogue tables are already beyond the state of the art, they can be further developed for subsequent releases. First, the same procedures can be extended to the Northern Galactic Hemisphere region ($b>20$~deg, $dec<0$). Second, additional information can be added through near-UV and far-UV detections in GALEX, which we chose not to include here due to its rather shallow limits especially for extragalactic objects. Third, increased spatial resolution can be obtained through the ESA-\textit{Gaia} \citep{GaiaDR2} and Hubble Source Catalog \citep{Whit16}. In this first release, we have preferred not to include \textit{Gaia} and \textit{HST} cross-matches for two main reasons: the \textit{HST} catalogue has sparse waveband and positional coverage; and the \textit{Gaia} catalogue completeness towards fainter magnitudes or small object-separations is poor, at present. Dedicated cross-matching with \textit{Gaia} and \textit{HST}, as ancillary tables, will be part of future releases.

\pagebreak

 \begin{acknowledgements}
CS has received funding from the European Union's Horizon 2020 research and innovation programme under the Marie Sk{\l}odowska-Curie actions grant agreement No 664931. AA is supported by a grant from VILLUM FONDEN (project number 16599). This project is funded by the Danish council for independent research under the project ``Fundamentals of Dark Matter Structures'', DFF - 6108-00470.\\
Based on observations obtained as part of the VISTA Hemisphere Survey, ESO Progam, 179.A-2010 (PI: McMahon).\\ 
This publication has made use of data from the VIKING survey from VISTA at the ESO Paranal Observatory, programme ID 179.A-2004. Data processing has been contributed by the VISTA Data Flow System at CASU, Cambridge and WFAU, Edinburgh.\\
This project used public optical archival data from the Dark Energy Survey (DES), the SkyMapper Southern Sky Survey (SM), the the Panoramic Survey Telescope \& Rapid Response System DR1 (PanSTARRS1), Sloan Digital Sky Survey IV (SDSS DR14) and the The 6dF Galaxy Survey (6dFGS). \\
This publication makes also use of data products from the Wide-field Infrared Survey Explorer, which is a joint project of the University of California, Los Angeles, and the Jet Propulsion Laboratory/California Institute of Technology, funded by the National Aeronautics and Space Administration. 

This research has made use of the SIMBAD database, operated at CDS, Strasbourg, France; the VizieR catalogue access tool, CDS,
 Strasbourg, France (DOI : 10.26093/cds/vizier); of the the NASA/IPAC Infrared Science Archive, which is operated by the Jet Propulsion Laboratory, California Institute of Technology, under contract with the National Aeronautics and Space Administration; and the NOAO Data Lab. NOAO is operated by the Association of Universities for Research in Astronomy (AURA), Inc. under a cooperative agreement with the National Science Foundation.

\end{acknowledgements}

%-------------------------------------------------------------------

\end{document}